\documentclass[aip,pop,reprint,amsmath,amssymb,nofootinbib]{revtex4-1}

\usepackage{bm}
\usepackage{soul}   
\usepackage{url}
\usepackage{varioref}
\usepackage{graphics}
\usepackage{graphicx}
\usepackage{appendix}
\usepackage{color}
\usepackage{multirow}
\usepackage{verbatim}
\usepackage{colortbl}
\usepackage{cancel}
\usepackage{times}
\usepackage{multimedia}
\usepackage{color}
\usepackage{enumerate}
\usepackage{sidecap}
\usepackage{dcolumn}
\usepackage[usenames,dvipsnames,svgnames,table]{xcolor}
\usepackage[compact]{titlesec}
\usepackage{wrapfig}
\usepackage{nth}
\newcommand {\f} {\frac}

\newcommand {\mf} {\mathbf}
\newcommand {\be} {\begin{equation}}
\newcommand {\ba} {\begin{eqnarrray}}
\newcommand {\ee} {\end{equation}}
\newcommand {\ea} {\end{eqnarray}}
\newcommand{\ts}{\textsuperscript}

\def\th{\theta}

\def\part{\partial}

\def\curl{\nabla\times}

\def\Wci{\Omega_{ci}}

\newcommand{\bmin}{\begin{minipage}{0.495\textwidth}}
\newcommand{\emin}{\end{minipage}}

\def\Talf{\tau_{A}}

\def\mrat{m_i/m_e}

\def\Be{\beta_e}

\def\Dp{\Delta'}
\def\deg{^{\circ}}
\def\db{d_{\beta}}
\def\cb{c_{\beta}}
\def\By0{B_y^{(0)}}
\newcommand{\pt}[1]{\frac{\partial#1}{\partial t}}
\newcommand{\vvec}{\mathbf{v}}
\newcommand{\Bvec}{\mathbf{B}}
\newcommand{\Evec}{\mathbf{E}}
\newcommand{\Jvec}{\mathbf{J}}
\newcommand{\Avec}{\mathbf{A}}

\def\J0p{J'_{y0}}
\def\j0p{j'_{y0}}
\def\psit{\tilde{\psi}}

\titlespacing{\section}{0pt}{3mm}{2mm}
\titlespacing{\subsection}{0pt}{2mm}{2mm}
\titlespacing{\subsubsection}{0pt}{2mm}{2mm}

\begin{document}
   \title{A Two-Fluid Study of Oblique Tearing Modes in a Force-Free Current Sheet}
   \author{Cihan Akcay}
   \thanks{Electronic mail: akcay@lanl.gov}
   \author{William Daughton}
   \affiliation{Los Alamos National Laboratory, Los Alamos, New Mexico 87545}
    \author{Vyacheslav S. Lukin} 
   \affiliation{National Science Foundation, Greenbelt, Maryland 20771 }
   \thanks{Any opinion, findings, and conclusions or recommendations expressed in this material are those of the authors and do not reflect the view of the National Science Foundation.}
   \author{Yi-Hsin Liu} 
   \affiliation{NASA Goddard Space Flight Center, Arlington, Virginia 22230}

\begin{abstract}
Kinetic simulations have demonstrated that three-dimensional reconnection in collisionless regimes proceeds through the formation and interaction of magnetic flux ropes, which are generated due to the growth of tearing instabilities at multiple resonance surfaces. 
Since kinetic simulations are intrinsically expensive, it is desirable to explore the feasibility of reduced two-fluid models to capture this complex evolution, particularly, in the strong guide field regime, where two-fluid models are better justified.
With this goal in mind, this paper compares the evolution of the collisionless tearing instability in a force-free current sheet with a two-fluid model and fully kinetic simulations. 
Our results indicate that the most unstable modes are oblique for guide fields larger than the reconnecting field, in agreement with the kinetic results. 
The standard two-fluid tearing theory is extended to address the tearing instability at oblique angles.
The resulting theory yields a flat oblique spectrum and underestimates the growth of oblique modes in a similar manner to kinetic theory relative to kinetic simulations. 
 
\end{abstract}
\maketitle
\section{Introduction} 
\label{sec:intro}
Magnetic reconnection\cite{PandF2000,Biskamp2000} is a fundamental process that rapidly converts plasma magnetic energy to kinetic energy by breaking and then reconnecting magnetic field lines 
embedded in the plasma. 
It is responsible for particle transport and energization observed in the solar corona\cite{Mikic1988,Forbes2006}, magnetosphere \cite{Pasch1979,Sonnerup1981}, various other astrophysical phenomena\cite{Uzden2006,Guo2014}, and many laboratory plasmas\cite{Hsu2000,Qin2001,Brown2002,Yamada2014}. 
While much insight has been gained from two-dimensional (2D) studies--see Ref.  \onlinecite{Ji_Daughton2011} and the references therein--a complete understanding of reconnection requires a fully three-dimensional (3D) treatment. 

Kinetic simulations have recently demonstrated that 3D reconnection in collisionless regimes proceeds through the formation and interaction of magnetic flux ropes, which are generated due to growth of tearing instabilities at multiple resonance surfaces\cite{Daughton2011,Liu2013}. 
The subsequent nonlinear interactions of these flux ropes lead to the self-generation of multi-scale structures and intermittent multi-fractal turbulence \cite{Leo2013}. 
In force-free current sheets, both kinetic theory and particle-in-cell (PIC) simulations suggest that the fastest growing tearing modes are associated with the resonance surfaces on the edge of the sheet, and these modes give rise to the most prominent flux ropes\cite{Liu2013}.   
These \textit{oblique} modes are completely suppressed in 2D studies of reconnection, which have dominated the literature. 
Thus, to study the realistic dynamics of reconnecting current layers, it is critical to move towards 3D simulations. 

Since kinetic simulations are intrinsically expensive, it is desirable to explore the feasibility of reduced two-fluid models to capture this complex evolution, particularly, in the strong guide field regimes, where two-fluid models are better justified. 
Previous fluid-modeling efforts in the collisionless regime primarily focused on Harris sheet geometries with reduced single or two-fluid MHD models. 
Hall MHD simulations with no guide field showed tearing instability over a narrow range of oblique angles with the maximum growth rate corresponding to the resonance layer at the center of the sheet\cite{Cao1991}. 
Two-field\cite{Borg2005,Grasso2007} and four-field\cite{Grasso2012} simulations of two-fluid MHD with a guide field reported the coalescence of current layers and formation of complex 3D structures including vorticity layers due to the combined action of drifting X-points and mutual attraction of parallel currents. 
These vorticity layers were found to lead to the formation of secondary instabilities of the Kelvin-Helmholtz type in both 2D and 3D.
Structures on the order of the electron inertial-scales were observed.  
Recently, it was shown that the most unstable plasmoid mode in the constant-$\psi$ regime of reduced MHD is an oblique mode\cite{Baalrud2012}. 

This paper investigates the dynamics of the collisionless tearing mode in a force-free current sheet in the large guide field limit, using a two-fluid description with an ion-to-electron mass ratio, $\mrat=100$. 
The numerical implementation is carried out within the HiFi multi-fluid modeling framework\cite{Lukin2008}. 
The use of the full set of two-fluid equations provides a natural extension of the Hall and reduced MHD models of Refs. \onlinecite{Cao1991} and \onlinecite{Baalrud2012}, and generalizes the two and four-field models of Refs. \onlinecite{Borg2005} and \onlinecite{Grasso2012}. 
Analytical work that extends two-fluid tearing theory to oblique modes is presented. 
Oblique tearing simulations are run in 2D by rotating the equilibrium to select the resonant surface of a single oblique mode. 
A guide field with the uniform component, $b_g$, as large as ten times the in-plane field ($b_g=1-10$) is employed. 
This range is relevant for many astrophysical\cite{Kivel1995} and laboratory plasmas \cite{Gekel2014,Wesson2011,Intrator2013,ITER2013}. 
The linear growth rates are compared to those from kinetic theory and fully kinetic PIC simulations\cite{Liu2013}, as well as two-fluid theory. 
Our results indicate that the most unstable (fastest growing) modes are oblique for $b_g\ge1$, in agreement with Ref. \onlinecite{Liu2013}. 
For $b_g>>1$, the peak oblique growth rate significantly exceeds that of the mode whose resonance layer lies at the center of the current sheet (referred to as the symmetric mode henceforth, since the tearing eigenfunction is symmetric at this location). 
Two-fluid theory produces a flatter oblique spectrum and underestimates the oblique tearing growth rates relative to the non-oblique mode, in a similar manner to how kinetic theory compares with kinetic simulations. 

While our primary interest in this study is the collisionless limit, it is necessary to include   dissipative effects in the two-fluid model for reasons of numerical stability and convergence. 
The dissipative term that influences the tearing instability is hyperresistivity ($\eta_H$) that enters the generalized Ohm's law as an artificial electron viscosity term (see sections \ref{sec:2fl_th} and \ref{sec:2fl_model}). 
Since our goal is to compare with collisionless PIC simulations, we scan $\eta_H$ for each $b_g$ until convergence of linear growth rates is achieved, which indicates the transition into the collisionless regime. 
By incorporating $\eta_H$ in the extended two-fluid tearing theory, we provide an analytic expression in terms of the equilibrium quantities for the critical $\eta_H$ below which collisionless tearing takes places. 
The resulting theoretical prediction for this transition is consistent with the two-fluid simulations, justifying the validity of our conclusions in the collisionless limit. 
This result is important because it allows us to select an $\eta_H$ sufficiently small in order to compare with collisionless kinetic simulations. 
The scans over $\eta_H$ also indicate that the peak growth rate shifts from an oblique mode to the symmetric mode when $\eta_H$ is sufficiently high, suggesting a stronger suppression of oblique modes with hyperresistivity.  

This article is organized as follows. 
Section \ref{sec:oblq_th} is an overview of the collisionless theory of oblique tearing as presented in Ref. \onlinecite{Liu2013}. 
We present results from the oblique two-fluid tearing theory as well as the incorporation of hyperresistive dissipation into two-fluid theory in Section \ref{sec:2fl_th}. 
The details of the derivation are contained in the Appendix. 
Section \ref{sec:2fl_model} introduces the two-fluid equations and some details of the implementation within the HiFi multi-fluid modeling framework\cite{Lukin2008}. 
Section \ref{sec:2Dtear1} contains the results from 2D two-fluid simulations of oblique tearing and comparisons to kinetic simulations and theory as well as two-fluid theory. 
The main conclusions of this work and additional discussion are presented in Section \ref{sec:discuss}. 

\section{Oblique Tearing Theory}
\label{sec:theory}
\subsection{Kinetic Theory of Collisionless Tearing}
\label{sec:oblq_th}
This section reviews some key results from Refs. \cite{Liu2013,Baalrud2012}, which are critical for the comparisons shown in this paper. 
The equilibrium is a force-free ($\Jvec^{(0)}\times\Bvec^{(0)}=0$) current sheet with thickness $2\lambda$. 
{$\Bvec^{(0)}=B_0 \left[\tanh(z/\lambda)\hat{x}+\sqrt{b_g^2+sech^2(z/\lambda)}\hat{y}\right]$ is the equilibrium magnetic field with magnitude  $|\Bvec^{(0)}|=B_0\sqrt{1+b_g^2}$ and
$\Jvec^{(0)}=\nabla\times\Bvec^{(0)}/\mu_0$ is the equilibrium current density. 
The electron $\beta$ is defined in terms of the equilibrium quantities as
$\beta_e=2\mu_0 n T_e/[B_0^2(1+b_g^2)]$, where $T_e$ is the electron temperature and $n$ is ion/electron plasma density. 
For all the simulation work presented here, initial ion and electron temperatures are the same: $T_i=T_e$, which results in $\beta=2\beta_e=2\beta_i$.  

An oblique mode is a general tearing perturbation with both in and out-of-plane components, $\mf{k}=k_x\hat{x}+k_y\hat{y}$ with obliquity $\th\equiv \tan^{-1}(k_y/k_x)$. 
Such a perturbation (with $k_y\ne0$) shifts the location of the resonance layer ($\mf{k}\cdot\Bvec^{(0)}=0$) away from the center of the sheet ($z_s=0)$ to $z_s=-\lambda \tanh^{-1}[(1+b_g^2)^{1/2}\sin\th]$. 
This implies that sufficiently large 3D systems will permit numerous tearing modes growing at multiple resonance layers. 

Applying the asymptotic analysis of Ref. \onlinecite{Furth1963} for $k<<1$ and $k>>1$ to an oblique perturbation yields the following expression for the tearing drive  parameter\cite{Liu2013,Baalrud2012}:
\be
\label{eq:dp}
\Delta'\approx\f{2}{k\lambda^2}(1+b_g^2 \tan^2\th)-2k,
\ee
where $k^2=k_x^2+k_y^2$. 
Standard matching of inside and outside solutions\cite{Furth1963,Drake1977} yields the following linear growth rate from kinetic theory\cite{Liu2013} 
    \be
       \label{eq:gamma1}
     \f{\gamma}{kv_{Te}}\approx\f{d_e^2\Delta'}{2\sqrt{\pi}l_s \left[1+\sqrt{(m_e/m_i)(T_e/T_i)}\right]}\approx\f{d_e^2\Delta'}{2\sqrt{\pi}l_s},
    \ee
where $d_e=c/\omega_{pe}$ is the electron inertial length, $\omega_{pe}=(e^2 n_e/m_e\epsilon_0)^{1/2}$ is the plasma frequency, and $v_{Te}=\sqrt{2k_BT_e/m_e}$ is the electron thermal speed, and 
    \be
    \label{eq:Ls}
      1/l_s\equiv\f{1}{k}\left(\f{dk_{\parallel}}{dz}\right)_{z=z_s}=\f{\cos^2\th-b_g^2 \sin^2\th}{\lambda \cos\th(1+b_g^2)^{1/2}}
    \ee
is the magnetic shear length\cite{Drake1977}. 
The denominator for the final equality in Eq. (\ref{eq:gamma1}) has been simplified further as a result of $m_e/m_i<<1$ and $T_e=T_i$. 

Substitution of the $k$ and $\theta$ dependence into Eq. (\ref{eq:gamma1}) indicates that both $\th<\th_c\equiv \tan^{-1}(1/b_g)$ and $k\lambda\lesssim\sqrt{2}$ are required for an instability. 
Note $\Dp$ is a monotonically increasing function of $\th$ that is not bounded by $\th_c$. 
At $\th=\th_c$ the resonance surface moves to $z=\pm\infty$ while $l_s\rightarrow\infty$. 
For convenience we re-express the collisionless growth rate of Eq. (\ref{eq:gamma1}) in terms of the ion cyclotron frequency, $\Wci=e B_0/m_i$, defined with respect to the reconnecting field $B_0$: 
     \be
       \label{eq:gamma2}
        \f{\gamma}{\Wci}\approx\f{k d_e^3 (\mrat)\Delta'\sqrt{\Be(1+b_g^2)}}{2\sqrt{\pi}l_s }.
     \ee
\subsection{Extension of Two-fluid Tearing Theory to Oblique Modes}
\label{sec:2fl_th}
To treat oblique modes, we extend the collisionless two-fluid tearing theory applicable to current sheet systems with an arbitrary guide field and $\beta>2m_e/m_i$ as developed by Fitzpatrick and Porcelli\cite{Fitz2004}. 
Ref. \onlinecite{Mirnov2004} also independently worked on the same problem for a collisional plasma with a large guide field. 
More recent works on the topic, including a treatment of electron/ion gyroviscosity can be found in Refs. \cite{Fitz2007,Hoss2009,Fitz2010}. 
The present work does not consider gyroviscosity in the theory or simulations. 

The details of the derivation are presented in the Appendix where we show that the inner layer equations for an oblique tearing mode produce the same eigenvector equation as the well-known symmetric mode if one neglects the equilibrium current gradient ($ J_{y0}'\equiv\part J_{y0}/\part z =\part (\hat{y}\cdot \Jvec^{(0)})/\part z$) contribution to the inner layer.  
However, this effect may be significant as oblique resonance layers lie in regions where $J_{y0}'\ne0$. 
It was shown by Ref. \onlinecite{Bertin1982} that the inclusion of $J_{y0}'$ modifies the tearing growth rate in resistive MHD. 
The Appendix presents details on the incorporation of $J_{y0}'$ into the inner layer equations and the resulting change in the tearing eigenvector equation. 
A formal solution is left for a subsequent publication as the emphasis of this paper is on the comparison between two-fluid and kinetic simulations of oblique tearing. 
Here, we simply state the growth rate without the $J'_{y0}$ contribution in the small $\Dp$ regime, which is valid when $w\Dp<1$, where $w=d_e^2\Dp/(2\sqrt{\pi})$ is the width of the collisionless tearing layer according to Ref. \onlinecite{Drake1977}. 
Inserting the values from Table \ref{tab:Params1} into $w\Dp$ for $\mrat=100$, we obtain $w\Dp=0.1-0.7$ over $\th=0\deg-\th_{cr}$, indicating a regime of small-to-intermediate $\Dp$. 

After accounting for the normalizations used in Ref. \onlinecite{Fitz2004} (see the Appendix), we finally arrive at the expression for the oblique tearing growth rate for small $\Dp$ and $\beta<<1$ based on their Eq. (78): 
\be
\label{eq:2fl_gam}
\f{\gamma^{(2fl)}}{\Wci}=\f{kd_e^3(\mrat)\Dp \sqrt{\Be(1+b_g^2)}}{\sqrt{2}\pi l_s},
\ee
Comparing Eqs. (\ref{eq:gamma2}) and (\ref{eq:2fl_gam}) indicates that in the limit of small $\Dp$ and $\beta<<1$ the growth rate from two-fluid theory has the same parametric dependence as that of kinetic theory and differs from it by a numeric constant: $\gamma^{(2fl)}/\gamma^{Kin}=\sqrt{2/\pi}$. 

For an arbitrary $\Dp$, one must use Eq. (95) of Ref. \onlinecite{Fitz2004} or Eq. (73) of Ref.  \onlinecite{Mirnov2004} to calculate the growth rate. 
However, for the parameters considered in this manuscript both small and arbitrary$\Dp$-approaches yield the same growth rates. 

By incorporating hyperresistivity ($\eta_H$) into the theory, one can determine where the transition into the regime of collisionless tearing occurs. 
For $\eta_H\ne0$, the Fourier transformed form of the governing equation for the inner (tearing) layer (Eq. (71) of Ref. \onlinecite{Fitz2004}) changes to
\be
\label{eq:inner_eqn1}
\f{\part}{\part r}\left[\f{r^2}{1+r^2+r^4\bar{\eta}_H}\f{\part \bar{Z}}{\part r} \right]-Q^2(1+\cb^2 r^2)\bar{Z} = 0,
\ee
where $r=p d_e$ is the dimensionless momentum, $\bar{Z}$ is the Fourier transformation of the eigenfunction $Z$ ($Z$ is the perturbation in the guide field or plasma pressure due to the tearing), $Q=\gamma/(k\db)$ is the rescaled growth rate, and $\bar{\eta}_H=\eta_H/\gamma d_e^4$. 
The parameter $\bar{\eta}_H$ naturally arises if one uses a heuristic argument to replace the effective skin depth $\delta=(d_e^2+\eta/\gamma)^{1/2}$ of Ref. \onlinecite{Mirnov2004} with its hyperresistive counterpart defined as $\delta=(d_e^4+\eta_H/\gamma)^{1/4}=d_e(1+\bar{\eta}_H)^{1/4}$. 
Comparing the $r^2$ to $r^4\bar{\eta}_H$ term in the denominator of Eq. (\ref{eq:inner_eqn1}) reveals the dissipation scale: $l_H\equiv\sqrt{\eta_H/\gamma d_e^2}$. 
The transition into the collisionless regime occurs when $l_H\le w$, which yields the following critical hyperresistivity:
\be
\label{eq:etaH_crit}
\eta_H^{cr}=\gamma d_e^2 w^2=\f{\gamma \Dp^2 d_e^6}{4\pi}.
\ee
We label the regime where $ l_H\le w$ collisionless and $l_H> w$ collisional. 

One can further evaluate Eq. (\ref{eq:etaH_crit}) by inserting the definitions of $\Dp$, $l_s$, and $\gamma$ (Eqs. (\ref{eq:dp}), (\ref{eq:Ls}), and (\ref{eq:2fl_gam}), respectively). 
The resulting expression is strictly in terms of the known quantities such as $k$, $b_g$ and $\th$.  
Thus, for any hyperresistive two-fluid system, we can determine \textit{a priori} where the collisionless regimes occurs. 
This result is important because it allows us to select an $\eta_H$ sufficiently small in order to compare with collisionless kinetic simulations. 
\section{Description of the Computational Models} 
\label{sec:2fl_model}
We employ a two-fluid model of a fully ionized plasma comprising isothermal electrons and adiabatic ions. 
This is the minimal two-fluid model that still contains the physical effects relevant for tearing in the weakly collisional regimes. 
The following equations comprise the two-fluid system:
\begin{align}
     \label{eq:cont}
    \pt{n} +\nabla\cdot \left(n\mf{u}\right) = 0, \\
    \label{eq:CM}
    \pt{\left(n\mf{u}\right)} +
   \nabla\cdot[n(\vvec_i\vvec_i+\vvec_e\vvec_e)+p\bar{\mf{I}}+\pi_i+\pi_e]  = \Jvec\times\Bvec,  \\
      \label{eq:OhmsLaw}
   \Evec+\vvec_e\times\mf{B}+\f{m_e}{m_i}\left(\pt{\vvec_e}+\vvec_e\cdot\nabla\vvec_e\right) = -\f{1}{n}\nabla\cdot( p_e\bar{\mf{I}}+\pi_e)+\eta\Jvec,
   \\
   n\left[\pt{T_i} + \vvec_i\cdot\nabla T_i+(\Gamma_i-1)T_i\nabla\cdot\vvec_i\right]
    =  \pi_i:\nabla\vvec_i-\nabla\cdot\mf{q},
    \label{eq:Tempr}
\end{align}
where $\Bvec=\nabla\times\Avec$ and $\Evec=-\part\mf{A}/\part t$ are the magnetic and electric fields, $\Avec$ the vector potential, $\Jvec=\nabla\times\Bvec$ the current density, $n$ the plasma number density, $\vvec_{e,i}$ the electron and ion flows, $\mf{u}=\vvec_i+(m_e/m_i)\vvec_e$ the center-of-mass plasma flow velocity, and $T_{e,i}$ the electron and ion plasma temperatures. 
Both $\Bvec$ and $\Jvec$ are auxiliary variables calculated from $\Avec$. 
In writing the two-fluid MHD equations in this form, the Weyl gauge has been chosen by explicitly setting the electrostatic potential to zero and absorbing any electrostatic $\Evec$-field that may arise into $\Avec$\cite{Lukin2011}. 
Note that all the quantities in the above equations have been non-dimensionalized by appropriate combinations of the magnitude of the reconnecting field $B_0$, ion inertial length $d_i=d_e(\mrat)^{1/2}$, and a background density $n_0$. 
Thus, the simulation time is measured in terms of an Alfv\'{e}n transit time through one $d_i$: $\tau_a=d_i/v_a$ where $v_a$ is defined with respect the reconnecting field $B_0$. 
This choice also implies $\tau_a\Wci =1$. 

Periodic boundary conditions are imposed at the surfaces intercepted by $\Bvec^{(0)}$  ($\hat{n}=\hat{x},\hat{y}$). 
For the direction normal to $\Bvec^{(0)}$ ($\hat{n}=\hat{z}$), free-slip
hard wall boundary conditions are imposed on the ion/electron flow
with $\hat{n}\cdot\nabla(\hat{n}\times\vvec_{i,e})=\hat{n}\cdot\vvec_{i,e}=0$ and conducting boundary conditions are imposed on the EM fields: $\hat{n}\times\Evec=\hat{n}\times\part \Avec/\part t=0$. 
In addition, $\hat{n}\cdot\nabla T_i=0$.

Basic kinematic closures with spatially-uniform and constant coefficients are assumed for the electron and ion stress tensors $\pi_i\equiv-\mu\nabla\vvec_i$ and $\pi_e\equiv-\eta_H n\nabla\vvec_e$, and heat flux $\mf{q}=-\kappa\nabla T_i$ where $\mu$ and $\eta_H$ are kinematic ion and electron viscosity (hyperresistivity for electrons), $\kappa$ is the heat conductivity, and $\eta$ is the plasma resistivity. 
The chosen values are set low enough to provide necessary dissipation for numerical stability without causing too much diffusion. 
Note the assumption of isothermal electrons neglects the viscous heating of the electron fluid due to hyperresistivity. 
However, the resulting leak in the total energy is completely negligible for the values of $\eta_H$ employed in the present simulations.  

Eqs. (\ref{eq:cont})--(\ref{eq:Tempr}) along with Amp\`{e}re's law, $\curl\curl\Avec=\Jvec$, are solved in a slab geometry using the high order spectral element multi-fluid modeling framework HiFi\cite{Lukin2008}. 
Spatial discretization is implemented with spectral elements where both the number of elements (or cells) $(n_x,n_y,n_z)$ and order of the polynomial representation for the nodal/modal basis functions $n_p$ are specified. 
The effective grid resolution is the product of the two quantities $(N_x,N_y,N_z)=n_p\times(n_x,n_y,n_z)$.
The time-stepping algorithm is implemented with the second order backward differencing (BDF2) method\cite{Bank1985}. 

The fully kinetic simulations were performed with the particle-in-cell code VPIC\cite{Bowers2008}. 
As the kinetic model was described in Ref. \onlinecite{Liu2013}, we omit its details in this section. 
The specific simulation parameters for both models are presented in section \ref{sec:2Dtear1}. 
\section{Two-fluid Simulations of Oblique Tearing}
\label{sec:2Dtear1}
\subsection{Setup}
The analytic form of the equilibrium is given in section \ref{sec:oblq_th}. 
For each simulation, the equilibrium is rotated about $\hat{z}$ by a specific $\theta$ to single out the resonance surface of one particular oblique mode, reducing the problem to 2D. 
The simulation domain has dimensions $L_x\times L_z=(2\pi\times2\pi) d_i$ with $\lambda=0.5d_i$. 
The geometry admits modes with $k\lambda\ge 0.5$.  
Longer-wavelength modes are excluded because PIC simulations indicate the fastest growing tearing modes to have $k\lambda\approx0.5$. 
We verified with two additional two-fluid simulations with an $L_x$ twice and three times that of the original setup that $k\lambda\approx0.5$ is still the fastest even though longer-wavelength modes ($k\lambda <0.5$) are allowed. 

The physical parameters for the simulations match those from Ref. \onlinecite{Liu2013}. 
The chosen regime of $\Be\sim 0.01-0.2$ is relevant to the solar wind and corona, and planetary magnetosphere because of the weak magnetic shear angles it produces. 
The first three rows of Table \ref{tab:Params1} summarize the key physical parameters. 
The uniform component of the guide field is set to $b_g =1$, 2.5, 4, and 10. 
The scans at $b_g=1-4$ have $T_e=0.09$, yielding $\beta_e=0.09$, 0.025, and 0.01, respectively. 
The scan at $b_g=10$ has $T_e=0.9$ $(\Be=0.02)$ to ensure $\rho_s>d_i$. 
The ion-to-electron mass ratio $\mrat=100$ ($d_i/d_e=10$). 
The dissipation coefficients are set to $\mu=10^{-4}$, $\eta_H=10^{-7}$, $\kappa=10^{-4}$, and $\eta=0$. 
Partial convergence tests were conducted with $\kappa=10^{-6}$ and $10^{-5}$, $\mu=10^{-5}$ and $\mrat=400$. 
Two additional full scans at $\eta_H=2.2\times10^{-5}$ and $10^{-6}$ as well as partial scans at certain oblique angles with $\eta_H$ up to $10^{-3}$ were also run to chart the dependence on $\eta_H$. 
The kinetic simulations have $\omega_{pe}/\Omega_{ce}=2$ where $\Omega_{ce}=(m_i/m_e)\Wci$ is the electron gyro frequency. 
 \begin{table}[h]
    \begin{center}
    \caption{\label{tab:Params1} The key simulation parameters for the oblique tearing scans with $\mrat=100$. The ranges in the ratios with $w$ reflect the span of $\Dp$ over $\th=0\deg-\th_{cr}$. The spectrum of oblique angles ($\th$) unstable to tearing shrinks as $b_g$ increases. 
   Ratios of the width of the collisionless tearing layer $w$ to ion and electron gyro radii, $\rho_i$ and $\rho_e$, are included for reference.}
    \begin{tabular}{||c|| c c c c||}
          \hline 
           $\mf{b_g}$ &  \textbf{1} & \textbf{2.5} &  \textbf{4}&  \textbf{10}  \\
    \hline
     $\beta_e$  &  0.09 & 0.025  & 0.01 & 0.02 \\
     $\mf{\th_c}$   & $45\deg$  & $22\deg$ & $14\deg$ & 5.7$\deg$\\
     $w/d_e$   &  0.17-0.40 & 0.17-0.40 & 0.17-0.40 & 0.17-0.40  \\
     $w/\rho_i$   &  0.06-0.1 & 0.1-0.25 & 0.2-0.4 & 0.1-0.3  \\
     $w/\rho_e$   &  0.6-1 & 1-2.5 & 2-4 & 1-3 \\
    $\rho_e/d_e$ &  $0.30$ & $0.16$ & $0.10$ & 0.04 \\
  \hline\hline
   \end{tabular}
 \end{center}
 \end{table}

Tearing is seeded with a sum over sinusoidal harmonics: $\delta A_y=\delta \sum_{n=1}^{10} cos(2\pi n x/L_x)cos(\pi z/L_z)$, spanning a range of $k \lambda=1/2\rightarrow 5$ to prevent biasing any particular mode. 
For the 2D simulations featured in this section, only the perturbed quantities are evolved while the equilibrium fields are kept static in time. 
This assures an accurate characterization of the effect of dissipation on the growth rates because it prevents the hyperresistive decay of the equilibrium gradients that feed the tearing instability. 
Typical resolutions employ a sufficiently high number of 3\ts{rd} or 4\ts{th} order elements with a non-uniformly distributed grid along $\hat{z}$ to produce 8--11 grid points per $d_e$ inside the current sheet for $\eta_H< 10^{-6}$ and 4--8 grid points per $d_e$ for $\eta_H\ge 10^{-6}$. 
\subsection{Results}
 \begin{figure*}
      \includegraphics[trim =0mm 0mm 0mm 0mm, clip=true,width=\textwidth]{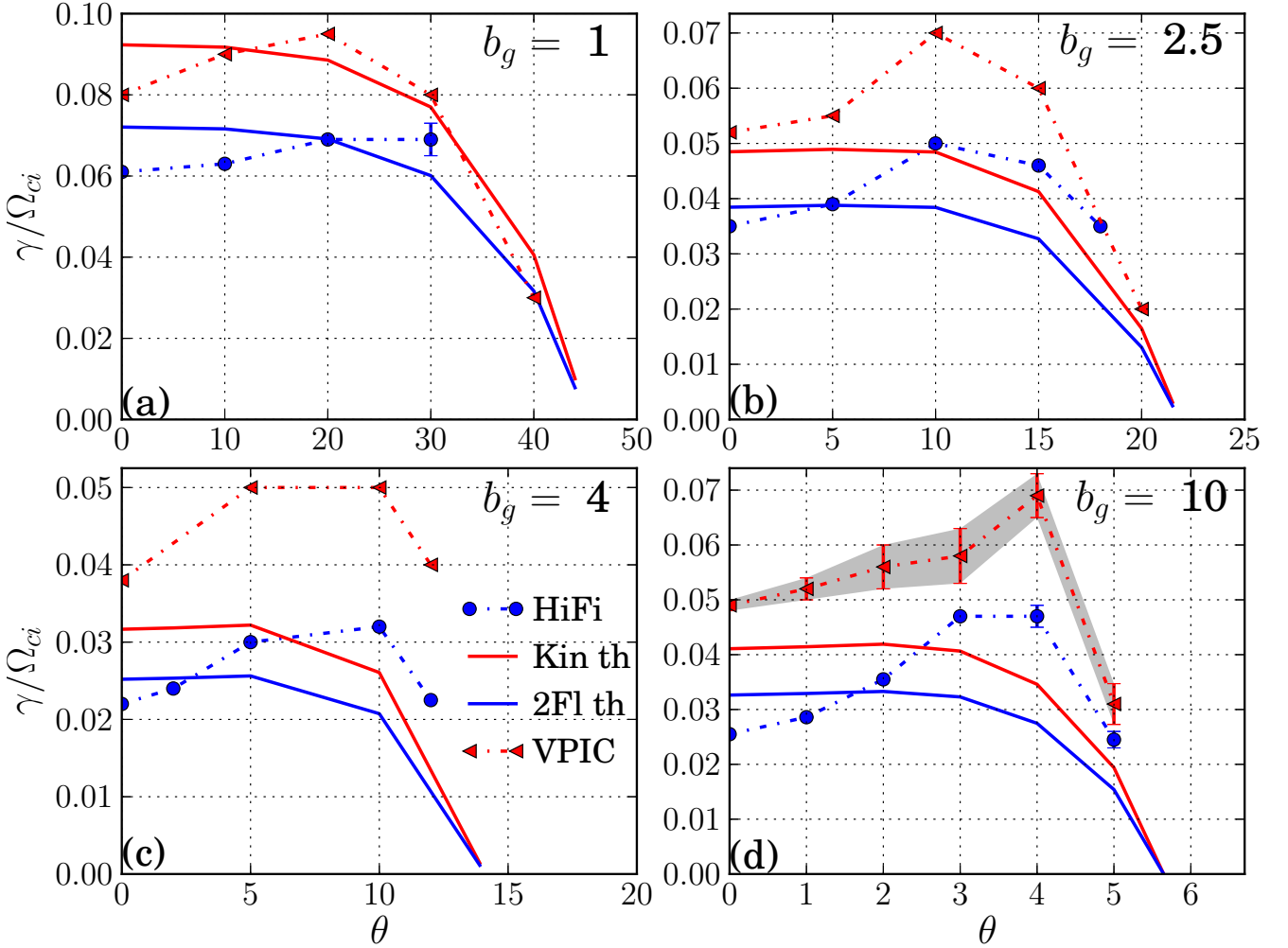}
        \caption{\label{fig:gamma_vs_theta_bg1_10} Normalized linear oblique tearing growth rates, $\gamma/\Wci$, as a function of the oblique angle $\th$ from HiFi two-fluid simulations (dashed blue with $\bigtriangleup$ ), VPIC kinetic simulations by Ref. \onlinecite{Liu2013} (dashed red with $\lhd$ ), two-fluid theory (solid blue), and linear kinetic theory (solid red) for $\mrat=100$ and (a) $b_g=1$, (b) $b_g=2.5$, (c) $b_g=4$, and (d) $b_g=10$. The range of the horizontal axis is decreased with increasing $b_g$ (decreasing $\th_c$) for clarity. The blue error bars indicate cases where the FFT and magnetic energy calculations of the growth rates differ by more than a few percent. The uncertainty in the VPIC data, shown for $b_g=10$, is represented by the shaded gray area.}
 \end{figure*}
Figure \ref{fig:gamma_vs_theta_bg1_10} shows the linear growth rates from the HiFi two-fluid simulations with $\eta_H=10^{-7}$ ( dashed blue with o), VPIC kinetic simulations\cite{Liu2013} (dashed red with $\lhd$), kinetic theory (solid red) and two-fluid theory (solid blue). 
The VPIC results\cite{Liu2013}are based on a series of 1D Fourier transformations of $B_z$ averaged over the entire thickness of the sheet ($d_i$). 
Recall $B_z=0$ initially and hence, it can only grow as a result of tearing. 
The growth rates from HiFi are computed in two ways: (1) by following the evolution of the magnetic energy component, $U_z=\int B_{z}^2 dV$, and (2) computing the FFT of $B_z$.
These two approaches for estimating the growth rate agree to better than 5\% except for a few cases with a large $\th$ that are plotted with an error bar to represent the standard deviation between the two calculations. 
The uncertainty in the VPIC data, shown for $b_g=10$, is represented by the shaded gray region in Fig. \ref{fig:gamma_vs_theta_bg1_10}d. 

 The figure shows that the most unstable modes from the two-fluid simulations are oblique with $k\lambda=0.5$, corresponding to a single X-point configuration, in agreement with VPIC simulations\cite{Liu2013} and reduced MHD\cite{Baalrud2012}. 
For $b_g>1$ oblique modes grow significantly faster than the parallel mode. 
Furthermore, the oblique modes become increasingly more unstable than the symmetric mode in the large guide field limit as evidenced by the trends in Figures \ref{fig:gamma_vs_theta_bg1_10}c and \ref{fig:gamma_vs_theta_bg1_10}d. 
At $b_g=10$ (Fig. \ref{fig:gamma_vs_theta_bg1_10}d), the peak growth rate (at $\th=3-4\deg$) is almost twice that at $\th=0\deg$ for both HiFi and VPIC. 
The peak growth rates from HiFi occur at similar locations to those from VPIC: $\th=20-30\deg$ for $b_g=1$, $\th=10\deg$ for $b_g=2.5$, $\th=5-10\deg$ for $b_g=4$, and $\th=3-4\deg$ for $b_g=10$.  
As predicted by the theory, HiFi growth rates are consistently lower than those from the VPIC kinetic simulations across the whole $\th$ spectrum. 
For $b_g\le2.5$, the offset between HiFi and kinetic simulations is  25--30\%, which is comparable to the 20\% offset between the theoretical traces. 
For $b_g=4$ and $10$ the growth rates from the two simulated models differ by as much as a factor of two for certain oblique angles. 
Both theories produce a flatter oblique spectrum and underestimate the oblique tearing growth rates. 
A possible reason for why this is the case in two-fluid theory is the exclusion of the finite equilibrium current gradient $J_{y0}'$ from the inner layer equations (see Ref. \onlinecite{Bertin1982} for a treatment of this effect in resistive MHD) .
An analysis of the modified inner layer equations as well as the resulting change in the tearing eigenvector equation is presented in the Appendix. 
There is a fair agreement between HiFi and two-fluid theory at $\th=0\deg$. 
Overall, the growth rates from the theory and simulations of the two models are within a factor of two of each other and the HiFi two-fluid simulations demonstrate that the fastest tearing modes in a force-free current sheet are oblique. 
\begin{figure}
   \begin{center}
        \includegraphics[trim =0mm 4mm 0mm 0mm, clip=true,width=0.5\textwidth]{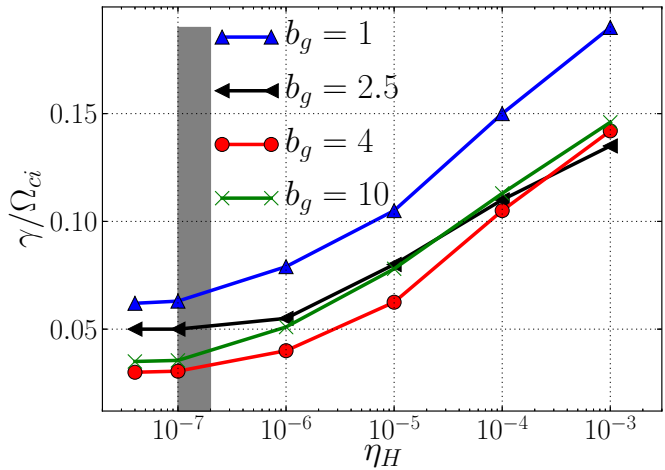}
   \end{center}
     \caption{ \label{fig:etaH_scan} The growth rates, $\gamma/\Wci$, as a function of the hyperresistivity $(\eta_H)$ for $b_g=1$, 2.5, 4, and 10 over the range $\eta_H=4\times10^{-8}-10^{-3}$. Growth rates converge for $\eta_H\le10^{-7}$, indicating the transition into the collisionless regime. The shaded region represents the theoretical prediction for where collisionless regime should occur based on Eq. (\ref{eq:etaH_crit}).}
\end{figure}

Sensitivity to hyperresistivity has been checked with scans over $\eta_H=4\times 10^{-8}-10^{-3}$ that spans collisionless to collisional regimes for $b_g=1-10$.  
For each value of $\eta_H$ and $b_g$, a single simulation with the oblique mode near or at the maximum growth rate was performed. 
The results are plotted in Figure \ref{fig:etaH_scan}.
The dashed blue, black, red, and green traces represent the linear growth rates as a function of $\eta_H$ for $b_g=1$, 2.5, 4, and 10, respectively. 
All four scans show a convergence of growth rates in the vicinity of $\eta_H=10^{-7}$, indicating a transition into the collisionless regime. 
Inserting the exact values of $\Dp$ and $\gamma$ into the theoretical prediction by Eq. (\ref{eq:etaH_crit}) yields $\eta_H^{cr}=1-2\times10^{-7}$, which agrees with the numerical results. 
In the collisional regime, the observed linear growth rates for the oblique modes vary as $\gamma\propto\eta_H^{0.13-1/6}$ while those for the symmetric mode $(\th=0\deg)$ vary as $\gamma\propto\eta_H^{1/4}$ (not shown here). 
Thus, in the collisional regime the dependence on hyperresistivity is weaker for oblique modes than the standard tearing mode. 
The 1/4-dependence at $\th=0\deg$ agrees with that reported by Ref. \onlinecite{Cai2010} for electron MHD tearing. 

The progression of the oblique spectrum from collisionless to collisional regimes for $b_g=2.5$ (top) and 4 (bottom) is shown in Figure \ref{fig:lin_rates_etaHscan}. 
The red, green, black, and blue traces correspond to $\eta_H=10^{-7}$, $10^{-6}$, $8\times10^{-6}$, and $2.2\times10^{-5}$, respectively. 
In both cases, as the dissipation is raised, the peaked spectrum flattens and the location of the peak growth rate increasingly shifts toward $\th=0\deg$ until $\gamma$ becomes a monotonically decreasing function of $\th$ at $\eta_H\gtrsim 10^{-5}$ (inside the collisional regime). 
This result is consistent with the above finding that oblique growth rates have a weaker dependence on $\eta_H$ inside the collisional regime. 
\begin{figure}
    \includegraphics[trim = 0mm 0mm 90mm 0mm, clip=true,width=0.49\textwidth]{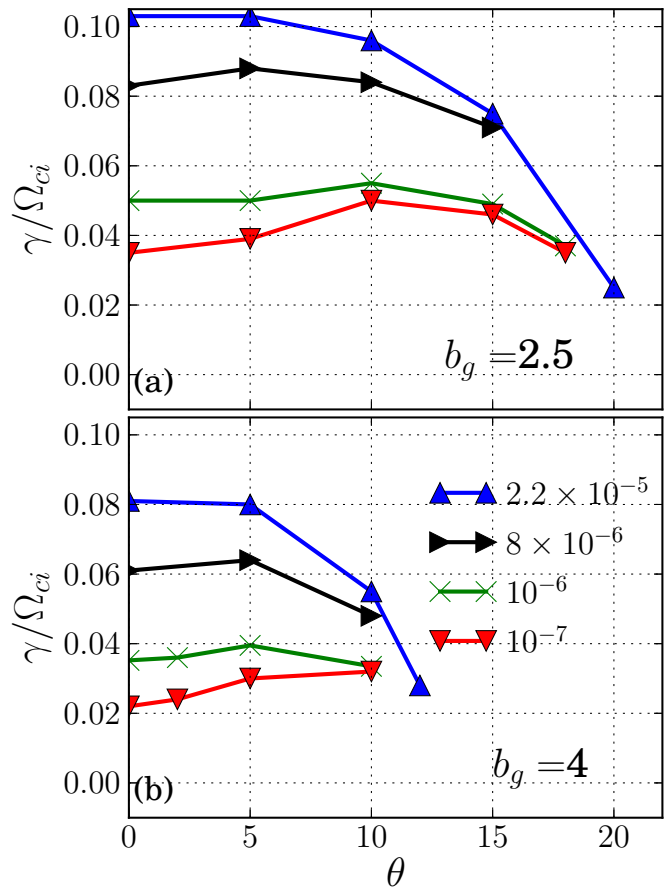} 
\caption{\label{fig:lin_rates_etaHscan} Linear growth rates ($\gamma/\Wci$) as a function of the obliquity ($\th$) for various values of $\eta_H$ at (a) $b_g=2.5$ and (b) $b_g=4$. As $\eta_H$ is raised from the collisionless value ($10^{-7}$), the peaked spectrum flattens and the location of the peak growth rate increasingly shifts toward $\th=0\deg$. At $\eta_H=2.2\times 10^{-5}$ (inside the collisional regime) the growth rate peaks at $\th=0\deg$ and monotonically decreases as a function of $\th$}. 
\end{figure}

In Figure \ref{fig:OhmsLaw_mime100} we plot the field-aligned (parallel) component of each term in Ohm's law (Eq. (\ref{eq:OhmsLaw})) along $\hat{z}$ to determine which terms generate the non-ideal (reconnection) electric field $E_{\parallel}$. 
Shown are profiles from $b_g=2.5$, $\th=5\deg$ (top panel) and $b_g=4$, $\th=2\deg$ (bottom panel) through the X point, spanning a distance a little over a $d_i$ at a time when $B_z\sim10^{-5}$. 
The thick solid vertical black line marks the location of the resonance surface for each case and the dashed vertical black lines demarcate the equilibrium current layer. 
The horizontal black trace with the upright triangles represents the sum of all terms, which must add up to zero for momentum conservation. 
The small bumps that appear in this trace are an artifact of the post-processing and have no physical meaning. 
The multiplicative factors appearing in front of each term  in Eq. (\ref{eq:OhmsLaw}) are omitted from the legend for brevity and the magnitudes are normalized to unity. 
The behavior is qualitatively the same for both cases and is representative of the parallel electron force balance for all $b_g$ and $\th$. 
The non-ideal electric field (blue) at the X-point is mainly supported by $\partial v_{e\parallel}/\partial t$ (red). 
As the mode grows the hyperresistive contribution embedded in $\nabla\cdot P_{e\parallel}/n$ (purple) also begins to support $E_{\parallel}$. 
Outside the resonance layer the contribution from the time derivative component of the electron inertia decreases while that from the hyperresistivity increases. 
The advective piece, $\mathbf{v_e}\cdot\nabla v_{e\parallel}$ (yellow), also registers some activity in this region. 
Beyond this zone, several $d_e$'s away from the X-point $E_{\parallel}$ vanishes.
 \begin{figure}
    \includegraphics[width=0.48\textwidth]{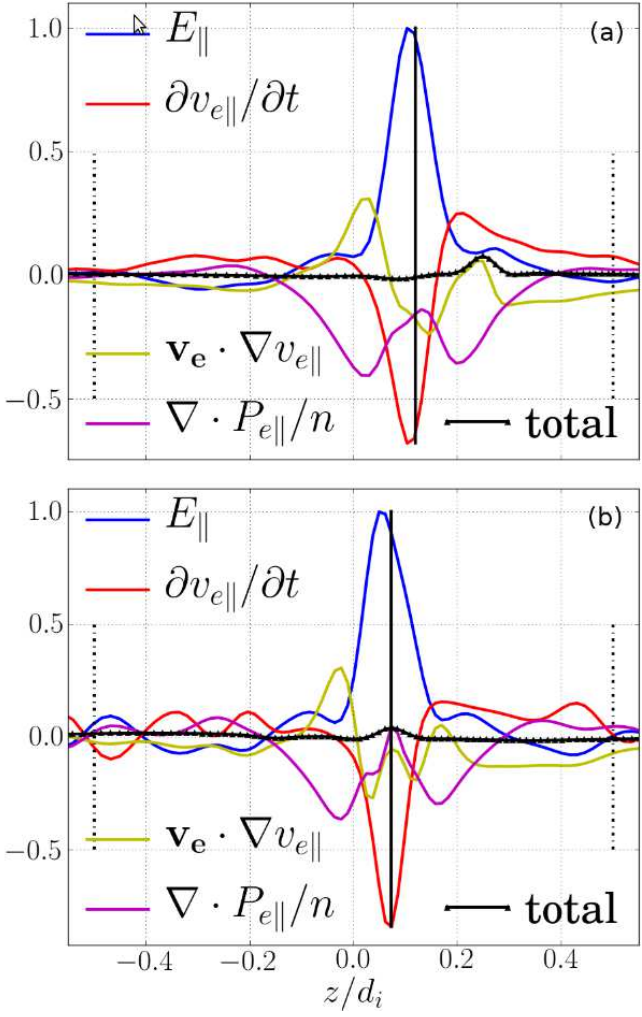}         
    \caption{\label{fig:OhmsLaw_mime100} Parallel (field-aligned) electron force balance. Shown are profiles of each term in the generalized Ohm's law, Eq. (\ref{eq:OhmsLaw}), for (a) $b_g=2.5$, $\th=5\deg$ and (b) $b_g=4$, $\th=2\deg$ at a time when $B_z\sim 10^{-5}$. The blue line corresponds to the non-ideal parallel electric field $E_{\parallel}$, the red and yellow lines to electron inertia terms $\partial v_{e\parallel}/\partial t$ and $\mathbf{v_e}\cdot\nabla v_{e\parallel}$, the purple line to the electron pressure tensor $\nabla\cdot P_{e\parallel}/n$, and lastly the black line with $\triangle$ to the sum of all the terms. Momentum conservation requires a sum of zero. The multiplicative factors appearing in front of each term in Eq. (\ref{eq:OhmsLaw}) are omitted from the legend. The time derivative of the electron inertia ($\partial v_{e\parallel}/\partial t$) is the primary generator of $E_{\parallel}$ due to tearing at the time shown in the figures.}
 \end{figure}

\section{Summary}
\label{sec:discuss}
We have performed two-fluid simulations of the oblique tearing instability in a force-free current sheet equilibrium with a guide field up to ten times the in-plane reconnecting field. 
By rotating the equilibrium, we single out a particular oblique tearing mode and investigate its linear evolution in 2D. 
Within the collisionless regimes, our simulations show the oblique tearing modes to be most the unstable for large guide fields ($b_g\ge1$), which agrees with the kinetic simulations\cite{Liu2013}. 
This trend is more pronounced at stronger guide fields. 
For example, with $b_g=10$ the two-fluid simulations indicate that the oblique modes with a resonance layer on the edge of the sheet grow nearly twice as fast as the mode with a resonance layer in the center of the sheet (symmetric mode). 
Two-fluid growth rates are consistently lower than their kinetic counterparts across the whole $\th$ spectrum, which agrees with the theoretical trend (see below). 

In the collisional regime, as the hyperresistive dissipation is raised, the peak growth rate shifts further toward the symmetric mode ($\th=0\deg$). 
When the dissipation scale $l_H$ exceeds $d_e$, the oblique spectrum becomes a monotonically decreasing function of obliquity $\th$ with the maximum at $\th=0\deg$. 
This is consistent with our finding that the oblique tearing modes grow at a slower rate (weaker $\eta_H$-dependence) than the symmetric mode as $\eta_H$ is raised inside the collisional regime. 

We have complemented the numerical effort by extending the two-fluid theory of tearing\cite{Fitz2004,Mirnov2004} for the symmetric mode to oblique modes. 
In the limit of $b_g>>1$ and a negligible gradient in both the equilibrium current $J_{y0}$ and guide field $B_y^{(0)}\equiv\hat{y}\cdot\Bvec^{(0)}$, the equations that describe the singular layer reduce to the same linearized form as those for the collisionless symmetric tearing mode. 
The resulting two-fluid theory differs from the kinetic theory only by $\sqrt{2/\pi}$ (~20\%). 
This difference is comparable to the offset observed between two-fluid and kinetic simulations at $b_g=1-2.5$ and $b_g=10$. 
Both theories produce a flatter oblique spectrum and underestimate the oblique tearing growth rates.
For non-zero gradient in $\By0$, the change in the growth rates can be estimated based on Ref. \onlinecite{Cai2009}.  
For $b_g>1$, growth rates should be enhanced by a small correction: $\sim1+b_g^{-2}$, amounting to an increase of 16\%, 6\%, and 1\% for $b_g=2.5$, 4, and 10, respectively. 

While neglecting the gradient in $\By0$ is a fairly good approximation for $b_g>>1$, the effect of the finite equilibrium current gradient, $J'_{y0}$, could be significant. 
This effect results in a correction to the tearing growth rate in resistive MHD\cite{Bertin1982}. 
Since $J_{y0}'$ feeds the tearing instability, it is possible a more pronounced oblique spectrum could emerge upon incorporating it back into the inner layer equations. 
As shown in the Appendix, the additional term due to $J'_{y0}$ is significant only for modes in the mid-obliquity range, drops sharply outside this range, and grows with $b_g$. 
These findings are consistent with the results of the two-fluid simulations. 
Also presented in the Appendix are the asymptotic limits of the modified eigenvector equation as a preview for the full treatment, which will be the topic of a follow-up paper. 
As for why kinetic theory predicts a flat oblique spectrum, this is a shortcoming of the boundary layer theory and is presently not well understood. 
 
Our simulations demonstrate that a two-fluid model captures the essential linear stage of the 3D tearing instability in a force-free current sheet equilibrium.
This provides a path forward for the continuation of this work where we aim to investigate the influence of line-tied boundary conditions on the nonlinear reconnection dynamics with both kinetic and two-fluid descriptions and compare the results from the two models. 

\section*{Acknowledgements}
We thank John Finn and Adam Stanier for valuable discussions and the anonymous referees for their suggestions. 
CA also thanks his late co-mentor Thomas Intrator for making his postdoctoral appointment at the Los Alamos National Laboratory (LANL) possible. 
This research was supported by funding from the Office of Fusion Energy Sciences, from the UCOP program from the University of California under Grant No. 12-LR-237124, and the LANL Center for Nonlinear Studies. 
VSL acknowledges support from the National Science Foundation. 
We used the resources of the LANL Institutional Computing Program supported by DOE/NNSA under Contract No. DE-AC52-06NA25936 and those of the National Energy Research Scientific Computing Center, a DOE Office of Science User Facility supported under Contract No. DE-AC02-05CH11231.
 

\appendix
\section{Appendix: Two-FLuid Theory of Oblique Tearing}
Fitzpatrick and Porcelli express the collisionless two-fluid system in terms of the four-field equations\cite{Aydemir1991}, which they solve, using the standard boundary layer theory by splitting the inner layer into two domains: an innermost region that extends from reconnection scales to electron scales ($z\lesssim d_e$) and a broader region that is matched to the usual MHD solution at $z>> d_e$. 
They carry out their analysis for small $\Dp$ ($w\Dp<1$), large $\Dp$ ($w\Dp>>1$), and arbitrary $\Dp$ where $w=d_e^2\Dp/(2\sqrt{\pi})$ is the width of the collisionless tearing layer according to Drake and Lee\cite{Drake1977}. 

In the presence of non-uniform perturbations along the guide field ($\part/ \part y\ne0$) and $\eta_H\ne0$, the dimensionless four-field equations of Ref. \onlinecite{Fitz2004} take on the following more generalized form:
\begin{align}
 \begin{split}\label{eq:psi_e}
\f{\part \psi_e}{\part t }={}& [\phi,\psi_e]+\db[\psi,Z]-\db b_g\f{\part Z}{\part y}+\f{\part \Phi_{es}}{\part y} -\eta_H\nabla^4 \psi,
\end{split}\\
\begin{split} \label{eq:Z_e}
\f{\part Z_e}{\part t }={}&[\phi,Z_e]+\cb[v_y,\psi]+\db[\nabla_{\perp}^2\psi,\psi]+\db b_g \f{\part \nabla_{\perp}^2\psi}{\part y}\\ 
                  &+\cb b_g \f{\part v_y}{\part y},
\end{split}\\
\label{eq:phi}
\f{\part \nabla_{\perp}^2\phi}{\part t }={}&[\phi,\nabla_{\perp}^2 \phi]+[\nabla_{\perp}^2\psi,\psi]+b_g\f{\part \nabla_{\perp}^2\psi}{\part y},\\
\label{eq:vz}
\f{\part v_y}{\part t }={}&[\phi,v_z]+\cb[Z,\psi]+\cb b_g\f{\part Z}{\part y},
\end{align}
where $Z$ is the perturbed component of the guide field scaled by $\cb\equiv\sqrt{\beta_e/(1+\beta_e)}$, $Z_e=Z-\cb^2 d_e^2\nabla_{\perp}^2 Z$, $\psi$ is the flux function such that $B_{\perp}=\nabla\psi\times\hat{y}$, $\psi_e=\psi-d_e^2\nabla_{\perp}^2\psi$, $v_y$ is the ion flow along the guide field, $\phi$ is the ion vorticity $v_{\perp}=\nabla\phi\times\hat{y}$, and  
$\Phi_{es}=-b_g\phi$ is the electrostatic potential, where $b_g>>1$ is assumed. 
$\perp$ denotes the direction perpendicular to the guide field ($\hat{y}$). 
$[A,B]=\nabla_{\perp} A\times  \nabla_{\perp} B\cdot \hat{y}$ is the usual Poisson bracket. 

The dissipationless $(\eta_H=0)$ form of the above equations was first derived by Grasso \textit{et. al.}\cite{Grasso2012}. 
Note our coordinate convention differs from that of Refs. \onlinecite{Grasso2012}, \onlinecite{Fitz2004}, \onlinecite{Mirnov2004}, and \onlinecite{Fitz2010} in that the guide field is along $\hat{y}$ instead of $\hat{z}$, the in-plane field is along $\hat{x}$ instead of $\hat{y}$, and the equilibrium gradients are along $\hat{z}$ instead of $\hat{x}$.
The coefficients $\cb$ and $\db=d_i\cb$ reduce to $\sqrt{\beta_e/2}$ and $\sqrt{\beta_e/2}d_i$, respectively for $\beta_e<<1 $. 
The additional factor of 2 appears because Ref. \onlinecite{Fitz2004} defines $\Be$ as $\bar{\beta_e}\equiv\Gamma_e P_e^{(0)}/B_0^2$, which is related to our $\beta_e$ in the following way: $\bar{\beta_e}=\Gamma_e\Be/2=\beta_e/2$ where $\Gamma_e=1$ for isothermal electrons.

Eqs. (\ref{eq:psi_e})-(\ref{eq:vz}) take on the following linearized form after applying the general form for an oblique perturbation $\tilde{\psi}(\mf{x})=\tilde{\psi}(z)e^{\gamma t + i (k_x x + k_y y)}$ and the usual inner layer ordering $\part/\part z> k_x,k_y$:
\begin{align}
\label{eq:lin_psi_e2}
g\left(\psit-d_e^2 \psit''+\f{\eta_H}{\gamma}\psit''''\right)&=i \f{\bar{z}}{l_s}(\phi-\db Z)-i\f{k_x}{k}d_e^2\j0p\phi, \\
\label{eq:lin_Z_e2}
g \left( Z-\cb^2 d_e^2 Z''+\f{\cb^2 \bar{z}^2 Z}{g^2 l_s^2}\right) &=i\db\left(\f{\bar{z}}{l_s}\psit''- \f{k_x}{k}\j0p\psit\right),
\end{align}
\begin{align}
\label{eq:lin_phi2}
g \phi''&= i \f{\bar{z}}{l_s}\psit''-i \f{k_x}{k}\j0p\psit,\\
\label{eq:lin_vz2}
g v_y &=i\cb\f{\bar{z}}{l_s}Z,
\end{align}
where $'$ denotes $\part/\part z$, $g=\gamma/k$, $\j0p\equiv (d_i^2\mu_0/B_0)\J0p$ is the non-dimensionalized equilibrium current gradient, and we note that when evaluating $[A,\psi] + b_g\part A/\part y=i(\mathbf{k}\cdot\Bvec^{(0)}/B_0)\tilde{A} + i k_x\psit A_0'$ in the vicinity of the resonance layer the first term can be approximated as $i(\mathbf{k}\cdot\Bvec^{(0)}/B_0)\tilde{A}|_{z=z_s}\simeq ik(z-z_s)\tilde{A}/l_s = i(k\bar{z}/l_s)\tilde{A}$ for any perturbed scalar field $A = \tilde{A} + A_0(z)$.

The fourth order term in Eqs. (\ref{eq:psi_e}) and (\ref{eq:lin_psi_e2}) arises due to hyperresistive dissipation.  
In the limit of zero dissipation, $\eta_H=0$, the inner layer equations for an oblique tearing mode as given by Eqs. (\ref{eq:lin_psi_e2})--(\ref{eq:lin_vz2}) differ from those for the symmetric tearing mode only in the terms proportional to $\j0p$ on the RHS of Eqs. (\ref{eq:lin_psi_e2})--(\ref{eq:lin_phi2}), which emerge out of the linearization of the Poisson brackets containing $\nabla^2\psi$. 
These terms are absent in the analysis for the symmetric mode for which $j_{y0}'=0$ at the resonance layer. 
However, as oblique modes arise in regions with strong equilibrium current gradients, this effect modifies the eigenvector equation. 
It was shown in Ref. \onlinecite{Bertin1982} that the inclusion of a non-zero $j_{y0}'$ modifies the tearing growth rate for resistive MHD. 

In the small $\Dp$ regime, the ion contribution to the inner layer equations is neglected\cite{Fitz2004}, meaning  all of the terms proportional to $\phi$ as well as the last term on the left hand side of Eq. (\ref{eq:lin_Z_e2}) ($\propto \cb^2 Z/g^2$) are excluded. 
This makes Eq. (\ref{eq:lin_phi2}) redundant. 
Note Eq. (\ref{eq:lin_vz2}) is decoupled from the system regardless of this approximation. 
Then, the only remaining term proportional to $\j0p$ appears on the RHS of Eq. (\ref{eq:lin_Z_e2}). 

If $\j0p=0$, the remaining equations reduce to the same eigenvector equation as in the case of symmetric tearing.
Hence, the rest of the analysis of Ref. \onlinecite{Fitz2004} directly applies. 
The oblique tearing growth rate for a small $\Dp$ and $\beta<<1$ as given by Eq. (78) of Ref.  \onlinecite{Fitz2004} is
\be
\label{Aeq:2fl_gam}
\gamma^{(2fl)}\Talf=\f{kd_e\Dp d_i\sqrt{\Be}}{\sqrt{2}\pi},
\ee
where $d_i\sqrt{\Be/2}$ is substituted $\db$ in Ref. \onlinecite{Fitz2004} and $\Talf=l_s/v_A$ is Ref. \onlinecite{Fitz2004}'s Alfv\'{e}n time defined with respect to the total magnetic field. 
As our simulation time is in terms of $\tau_a=d_i/v_a=(\Wci)^{-1}$, we carry out one additional step of arithmetic, using the relation $\tau_a/\Talf=(d_i/l_s)(v_A/v_a)=(d_i/l_s)\sqrt{1+b_g^2}$ to arrive at Eq. (\ref{eq:2fl_gam}). 

If $\j0p\ne0$, the resulting eigenvector equation in the Fourier domain becomes:
\be
\label{eq:eigenvector1}
r^2 \f{\part^2 \bar{Z}}{\part r^2}+ \left(\f{2r}{1+r^2}-if_{\th}\right)\f{\part \bar{Z}}{\part r}-Q^2(1+\cb^2r^2)(1+r^2)\bar{Z}=0,
\ee
where $f_{\th}=sin(2\th)cos(\th)(1+bg^2)d_e/(\lambda l_s)$ is the additional term that arises due to $\j0p\ne0$. 
Inserting Eq. (\ref{eq:Ls}) ($1/l_s$) into $f_{\th}$ and going to the limit of $b_g>>1$ ($\th<<1$) yields $f_{\th}=2b_g\th(1-b_g^2\th^2)(d_e/\lambda^2)$.
For $b_g=10$, this term has a maximum at $\th=3\deg$ (and at $\th=9\deg$ for $b_g=4$), is of order unity, and drops sharply outside the range $\th=2-4\deg$, consistent with the trend from the simulations (Figure \ref{fig:gamma_vs_theta_bg1_10}d). 
Thus, the modes in the mid-obliquity range should be most affected.  

The asymptotic limits of Eq. (\ref{eq:eigenvector1}) provide some insight. 
In the $r\rightarrow0$ limit (supra $d_e$ scales), the solution is a linear combination of confluent hypergeometric functions, in contrast to $\bar{Z}\sim \mbox{const}+ 1/r$ in the case of $\j0p=0$. 
The solution in the $r>>1$ limit, which corresponds to sub-$d_e$ scales that are more relevant for collisionless two-fluid tearing, remains unchanged. 
The full analysis is left for a follow-up paper as the focus of this paper is the comparison between the two-fluid and kinetic simulations of oblique tearing. 

\end{document}